\documentstyle[11pt]{article}
 
\newcommand{\beq}{\begin{equation}}
\newcommand{\eeq}{\end{equation}}
\newcommand{\bea}{\begin{eqnarray}}
\newcommand{\eea}{\end{eqnarray}}

\begin{document}
\title{Testing the Sprint Curve Model using the 150m Bailey--Johnson
Showdown}
 
\author{J.\ R.\ Mureika\thanks{newt@sumatra.usc.edu} ~
\\
{\it Department of Computer Science} \\
{\it University of Southern California} \\
{\it Los Angeles, California 90089~~USA} }

\maketitle
\vskip .25 cm
 
\noindent
{\footnotesize
{\bf Abstract} \\
Recently, a simple model was derived to account for a sprinter's energy
loss around a curve, based on previous sprint models for linear races.
This paper offers a quick test of the model's precision by comparing
split times from Donovan Bailey's 150m ``Challenge of Champions'' race
at Skydome on June 1st, 1997.  The discrepancy in the track configuration
which almost prompted Bailey to drop from the race is also addressed.} \\
\vskip .25 cm
 
In a highly controversial showdown at Skydome in Toronto on June 1st, 
Canadian sprinter Donovan Bailey proved that he is the 
``World's Fastest Man'' by defeating opponent Michael Johnson of the United 
States over an unconventional distance of 150m.  Bailey clocked a time
of 14.99s, a mere 0.02s off the ``official'' World Record of 14.97s,
held by Britain's Linford Christie.   

The showdown was set in motion
by each athlete's remarkable performance at the 1996 Olympic Games in
Atlanta, Georgia.  Bailey captured gold in the 100m with a World
Record time of 9.84s, while Johnson obliterated the existing 200m
World Record in a breathtaking 19.32s.   Following the latter of these
sprints, public opinion and media influence (not to mention American
nationalism) split the vote on who actually 
should be the bearer of the title ``World's Fastest Man'', a designation
traditionally reserved for the 100m champion.

A number of predictions were made in the buildup to this race as to the
value of the winning time.  Even contests were held, with prizes going
to the individual who could correctly guess the victor and his corresponding
victorious time.   The majority of these predictions were well under the
official 14.99s finish (see {\it e.g.} \cite{me2,tibs,globe}), and most
fell in the range of $14.70\sim 14.80$s.  So, to add to the disappointment of
Johnson's dropping out mid--way due to injury was  Bailey's curiously
``slow'' victory.

In this brief report, Bailey's performance will be discussed in light of
a simple model to calculate short sprint times for races which are partially
run off a curve \cite{me1}.  To quickly review the model and associated
history, one
must recall the underlying assumptions of the original model.  In the early 
1970s, J.\ Keller \cite{kel} proposed that the distance covered by a sprinter
over a given time could be expressed as the solution to the differential
equation

\beq
\dot{v}(t) = f(t) - \tau^{-1} v(t)~,
\label{hk}
\eeq
where $v(t)$ is the velocity at time $t$, $f(t)$ is a measure of 
force/unit mass
exerted by the athlete, and $\tau^{-1}$ is a decay term
(which crudely models fatigue--related factors).  From his analysis (an
optimization problem), Keller
determined that $f(t) = f \equiv constant$ for short sprints \cite{kel}.
The solution to (\ref{hk}) is found subject to the constraint $v(0)=0$, 
$f(t) \leq f$, and the resulting distance $d$ traveled in time $T$ is simply

\beq
d = \int_0^T dt \, v(t)~.
\label{dist}
\eeq
There is an additional equation which couples with (\ref{dist}) if the
race is longer than 291m, but this will not be addressed here.  The 
interested reader is directed to the citations in \cite{kel} for further
reading.

Prompted by the 150m showdown between Bailey and Johnson, R.\ Tibshirani
revised (\ref{hk}) in a statistical analysis based on the
Atlanta races which predicted Bailey would win by a margin of
$(0.02,0.19)$ seconds at a 95\% confidence level \cite{tibs}.  His adjustment 
to (\ref{hk}) consisted of the modification of $f(t)$, reasoning that it
is erroneous to set $f(t) = f= const.~\forall t$.  Rather, since a
sprinter must experience fatigue at {\em some} rate, 
it would make more sense to assume that $f(t) = f - ct$, for some $c > 0$.  

Tibshirani notes \cite{tibs}, however, that neither Keller's nor his 
model take into account the effects of the curve, {\it i.e.} their power 
of predictability drastically diminishes for races longer than 100m.  
In \cite{me1}, a simple term proportional to the centrifugal force felt
by the runner is introduced to Tibshirani's modification of (\ref{hk}).
Since ideally the two forces are normal to each other, the terms are
added vectorally.  The resulting equation of motion is

\beq
\dot{v}(t) = -\tau^{-1} v(t) + \sqrt{(f-ct)^2-
\lambda^2\frac{v(t)^4}{R^2}}~,
\label{mine3}
\eeq
for a track with curve of radius $R$.

The term $\lambda$ is introduced to account for the fact that a sprinter
does not feel the full value of the centrifugal force exerted on him/her
(by means of leaning into the turn, banked curves, different use of
leg muscles, etc...).  It is a simplified attempt to model a seemingly
non--trivial mechanism.  

The total distance of races run off the curve can be expressed as
$d = d_c + d_s$, where 

\bea
d_c & =& \int_0^{t_{1}} dt \, v_c(t), \nonumber \\
d_s & =& \int_{t_{1}}^T dt \, v_s(t)~, 
\label{curvedist}
\eea
with $v_c(t)$ the solution to Equation~(\ref{mine3}), and $v_s(t)$ the
velocity as expressed in the modified (\ref{hk}) (with $f(t) = f-ct$), 
subject to the boundary
condition $v_c(t_1) = v_s(t_1)$. Here, $t_1$ is the time 
required to run the curved portion of the race (distance $d_c$).

	Split times ({\it e.g.} the time $t_1$ to run the curve)
have never been recorded 
accurately until recently, so there has been up till now an unfortunate
lack of empirical data which could be used to make or break such a 
model.  Luckily, the 150m race held at Skydome was well--documented, and 
splits were obtained for the 50m and 100m marks; the former is on the curve,
the latter not.  The official splits \cite{torstar} for Donovan Bailey 
are given in Table~\ref{splits}.  If the model is an accurate representation 
of the physical process, a value of $\lambda^2$ can
be found to reproduce these times subject to the equations of motion.

The track used in Skydome was an unconventional configuration of
$d_c = d_s =~$ 75m (herafter denoted as 75m+75m), and was to have had 
a radius of curvature corresponding
to lanes 8 and 9 of a standard outdoor track \cite{fax}.  The corresponding
value of $R$ can be calculated as

\bea
R& = &\left(\frac{100}{\pi} + 1.25 (p - 1) \right)metres~, 
\label{radius}
\eea
for $p$ the lane number.
The form of (\ref{radius}) roughly results from the IAAF regulations 
governing curvature of outdoor tracks. The total curve length in 
lane $p=1$ (the smallest
radius of curvature) must be exactly 100m (the complete set of IAAF
rules and regulations can be obtained from \cite{iaafweb}).

For the equations of motion in (\ref{mine3}), the parameters obtained for
Bailey from a least square fit of his official Atlanta splits (see 
\cite{me1} for specific details) are used. These are 
$(f,\tau,c) = (7.96,1.72,0.156)$.  

The night prior to competition, Bailey threatened to drop out of the
race, citing that the track did not conform to the specifications agreed
upon in a signed contract.  In particular, he claimed that the curvature
of his lane corresponded more to lane 3 of an outdoor track rather than 
lane 8.  Additionally, he submitted that the curve was 10m longer than
anticipated, giving a 85m+65m configuration instead of 75m+75m \cite{bail1}.

Thus, (\ref{mine3}) is solved for the split distances 50m, 100m, as well
as the final 150m mark.  The model  in \cite{me1} does not account for 
the sprinter's reaction time, which must be added on to the resulting 
calculated times.  For Bailey, this was $t_{reac} =$+0.171s \cite{torstar}.

Tables~\ref{75ln3},~\ref{75ln8} give the splits for a 75m+75m configuration
total as run by Bailey in lanes 3 and 8, respectively.  
Tables~\ref{85ln3},~\ref{85ln8} present the same information for
an 85m+65m configuration.  Note that in this case, the splits would be
equal up to 75m, since (for equal lane assignments) the curve is of the same 
radius despite the fact
that it is longer.  So, splits are only given for 85m and beyond.
The sixth column lists the sum of difference of squares (a loose measure
of relative error from a small sample space)
$\Sigma^2 \equiv \sum_{i=50,100,150} \Delta_i^2/T_i^2$, where $\Delta_i = t_i - T_i$, for
$T_i$ the official splits of Table~\ref{splits} and $t_i$ the associated
model predictions.

The findings in \cite{me1} suggest that 
$\lambda^2$ could realistically assume a value between 0.50 and 0.80, 
so similar values are used here.  For interest's sake, possible 200m times 
for Bailey are extrapolated, to gauge whether or not he could be a viable
contender to Michael Johnson in the 200m, as stated in his post-race
interview.

Upon first inspection, the model \cite{me1} reproduces the official splits
and race time surprisingly well.
The smallest value of $\Sigma^2$ for each configuration is taken to be
the closest fit to the official race splits.  These are:

\begin{itemize}
\item{75m+75m, lane 3: $\lambda^2=0.50; \Sigma^2 = 8.48 \times 10^{-5}$}
\item{75m+75m, lane 8: $\lambda^2=0.80; \Sigma^2 = 7.90 \times 10^{-5}$}
\item{85m+65m, lane 3: $\lambda^2=0.50; \Sigma^2 = 2.71 \times 10^{-5}$}
\item{85m+65m, lane 8: $\lambda^2=0.70; \Sigma^2 = 4.22 \times 10^{-5}$}
\end{itemize}

Interestingly enough, the closest match above comes from the 85m+65m lane
3 configuration, which is the configuration allegedly used contrary to
the signed contracts.  However, it is obvious that a more precise choice
of $\lambda^2$ could easily yield a closer match for any of the 
configurations.  The different solutions arise from the readjustment of the 
ratio $\lambda/R$;
hence, to narrow down an ``exact'' solution (if indeed one exists), one
needs to analyze other race splits.  An interesting test of the model will
come at the end of June 1997, when Donovan Bailey will run a 150m race
at a Grand Prix meet in Sheffield, England (most likely on a 50m+100m
configuration; see \cite{me2} for Bailey's possible 150m times on such a
track).

  In \cite{me1}, some possible 200m times
were obtained which Bailey might be able to run under peak conditions.
It was found that $\lambda^2 \in [0.5, 0.7]$ yielded 200m times between
20.29s--20.59s, as run in lane 4 of an outdoor track,  which are
reproduced in Table~\ref{db200out}.  
These closely approximated Bailey's 1994 personal bests of 20.76s / 20.39s 
wind--assisted \cite{me1}).  

This season (1997), Bailey has clocked a 20.65s 200m \cite{canrank}, 
which agrees with the
cited range of $\lambda^2$. Assuming that the same value of $\lambda^2$ held 
for the 150m race in 
Skydome (between 0.7--0.8 \cite{me1}), then according to Table~\ref{75ln8},
then a possible configuration of the track would have been 75m+75m with
a curve radius equivalent to lane 8 (5.78s, 10.27s, 14.92s).
Accounting for the fact that Bailey was undoubtedly much more mentally and
physically prepared for the Skydome match, then it is likely that his
equivalent outdoor 200m time would drop.  The values $\lambda^2 = 0.6$ give
200m times between 20.29s -- 20.40s, which is a reasonable range for
Bailey to hit if he runs all--out.  It is not inconceivable that he
could run a low 20s race, despite his higher PB of 20.65s.  Both Frank
Fredricks of Namibia and Ato Boldon of Trinidad are comparable 100m runners,
and each has clocked a sub--20s 200m time (19.66s and 19.80s, respectively
\cite{iaafweb}).  Undoubtedly, it is high muscular endurance which allows 
them to do this.

For each case listed above, Bailey's extrapolated 200m times 
of Tables~\ref{75ln3}-\ref{85ln8} are generally less than for a standard 
outdoor track.  This is simply 
due to the fact that the longer curve outdoors (100m v.s. 75m) creates a 
larger drain on $f(t)$.  Whether or not such seemingly large time 
discrepancies are physically realizable, or are just a manifestation of the 
model, are unknown.  The proof is left to the sprinter.

\vskip .25 cm
 
\noindent
{\bf Acknowledgements}
Thanks to R.\ Mureika (U. of New Brunswick Department of Mathematics and
Statistics) for some useful suggestions, and to C.\ Georgevski (Director,
U. of Toronto Canadian High Performance Track and Field Centre) for making
it possible for me to observe first--hand the cited experiment in progress.

\pagebreak
\begin{table}
\begin{center}
{\begin{tabular}{|c|c|c|c|c|}\hline
Distance (m) & Split (s) \\ \hline
0 & 0.171 \\ 
50 & 5.74 \\
100 & 10.24 \\
150 & 14.99  \\ \hline
\end{tabular}}
\end{center}
\caption{Donovan Bailey's official splits for the Challenge of Champions
150m race at Skydome, Toronto, 01 June 1997 \cite{torstar}.}
\label{splits}
\end{table}

\begin{table}
\begin{center}
{\begin{tabular}{|c||c|c|c|c|c||c|}\hline
$\lambda^2$&$t_{50}$&$t_{75}$&$t_{100}$&$t_{150}$&$\Sigma^2$&$t_{200}$ \\ \hline
0.50&5.76 & 7.98 & 10.21 & 14.87 & $8.48\times 10^{-5}$  & 20.13 \\ \hline
0.60&5.79 & 8.03 & 10.28 & 14.94 & $1.02\times 10^{-4}$ & 20.22 \\ \hline
0.70&5.81 & 8.09 & 10.36 & 15.04 & $2.97\times 10^{-4}$   & 20.33 \\ \hline
0.80&5.84 & 8.13 & 10.41 & 15.10 & $6.34\times 10^{-4}$ & 20.41 \\ \hline
\end{tabular}}
\end{center}
\caption{Bailey's predicted splits as run in lane 3 for a 75m+75m track configuration.  All times include reaction time $t_{react} =$ +0.17s.}
\label{75ln3}
\end{table}

\begin{table}
\begin{center}
{\begin{tabular}{|c||c|c|c|c|c||c|}\hline
$\lambda^2$&$t_{50}$&$t_{85}$&$t_{100}$&$t_{150}$&$\Sigma^2$&$t_{200}$ \\ \hline
0.50&5.73 & 7.91 & 10.12 & 14.75 & $3.97\times 10^{-4}$  & 20.00 \\ \hline
0.60&5.75 & 7.95 & 10.17 & 14.82 & $1.79\times 10^{-4}$ & 20.07 \\ \hline
0.70&5.77 & 7.99 & 10.22 & 14.88 & $8.49\times 10^{-5}$ & 20.15 \\ \hline
0.80&5.78 & 8.02 & 10.27 & 14.92 & $7.90\times 10^{-5}$ & 20.20 \\ \hline
\end{tabular}}
\end{center}
\caption{Predicted splits as run in lane 8 for a 75m+75m track configuration.}
\label{75ln8}
\end{table}

\begin{table}
\begin{center}
{\begin{tabular}{|c||c|c|c|c|c||c|}\hline
$\lambda^2$&$t_{50}$&$t_{85}$&$t_{100}$&$t_{150}$&$\Sigma^2$&$t_{200}$ \\ \hline
0.50& 5.76 & 8.89 & 10.26 & 14.94 & $2.71\times 10^{-5}$ & 20.22 \\ \hline
0.60& 5.79 & 8.96 & 10.34 & 15.04 & $1.82\times 10^{-4}$ & 20.33 \\ \hline
0.70& 5.81 & 9.02 & 10.41 & 15.13 & $5.12\times 10^{-4}$ & 20.44 \\ \hline
0.80& 5.84 & 9.08 & 10.48 & 15.22 & $1.09\times 10^{-3}$ & 20.48 \\ \hline
\end{tabular}}
\end{center}
\caption{Predicted splits as run in lane 3 for an 85m+65m track configuration.}
\label{85ln3}
\end{table}

\begin{table}
\begin{center}
{\begin{tabular}{|c||c|c|c|c|c||c|}\hline
$\lambda^2$&$t_{50}$&$t_{85}$&$t_{100}$&$t_{150}$&$\Sigma^2$&$t_{200}$ \\ \hline
0.50& 5.73 & 8.80 & 10.15 & 14.80 & $2.41\times 10^{-4}$ & 20.06 \\ \hline
0.60& 5.75 & 8.84 & 10.20 & 14.87 & $8.24\times 10^{-5}$ & 20.13 \\ \hline
0.70& 5.77 & 8.89 & 10.26 & 14.94 & $4.22\times 10^{-5}$ & 20.22 \\ \hline
0.80& 5.78 & 8.93 & 10.30 & 15.00 & $8.34\times 10^{-5}$ & 20.29 \\ \hline
\end{tabular}}
\end{center}
\caption{Predicted splits as run in lane 8 for an 85m+65m track configuration.}
\label{85ln8}
\end{table}

\begin{table}
\begin{center}
{\begin{tabular}{|c||c c|c c| c| c| c|}\hline
$\lambda^2$&$t_{50}$&$v_{50}$&$t_{100}$&$v_{100}$&$t_{150}$&$t_{200}$&$t_{200}+0
.16$\\
\hline
0.36 & 5.55 & 11.60 & 9.98 & 10.85 & 14.69 & 19.96 & 20.12 \\ \hline
0.50 & 5.59 & 11.43 & 10.09 & 10.65 & 14.84 & 20.13 & 20.29 \\ \hline
0.60 & 5.61 & 11.31 & 10.16&10.51&14.93&20.24&20.40 \\ \hline
0.70 & 5.63 & 11.20 & 10.24 & 10.39 & 15.09 & 20.43 & 20.59 \\ \hline
\end{tabular}}
\end{center}
\caption{Bailey's predicted outdoor 200m times,
as run in lane 4.}
\label{db200out}
\end{table}
\end{document}